\documentclass[twoside,fleqn,a4paper]{article}
\usepackage{espcrc2}
\usepackage{graphicx}
\newcommand{\ACEP}[1]{\ensuremath{\left\langle #1 \right\rangle_{\bar\phi}}}
\title{Probing finite size effects in $\left( \lambda \Phi^4
\right)_4$ MonteCarlo calculations}
\author{A. Agodi \\ 
        G. Andronico\\
        Dipartimento di Fisica dell'Universit\'a di Catania\\
        INFN Sezione di Catania}
\begin{document}
\begin{abstract}
The Constrained Effective Potential (CEP) is known to be equivalent to
the usual Effective Potential (EP) in the infinite volume limit. We
have carried out MonteCarlo calculations based on the two different
definitions to get informations on finite size effects. We also
compared these calculations with those based on an Improved CEP (ICEP)
which takes into account the finite size of the lattice.  It turns out
that ICEP actually reduces the finite size effects which are more
visible near the vanishing of the external source.
\end{abstract}
\maketitle
\section{CEP and its properties}
The effective potential is defined as the Legendre transform of the
Schwinger function $W\left(\Omega,j\right)$ ($j$ constant external
source, $\Omega$ lattice size) 
\begin{equation}
\label{Gammadef}
  \Gamma \left( \Omega, \bar\phi \right) = \sup_j \left( j \bar\phi -
    W \left( \Omega , j \right) \right)
\end{equation}
with
\begin{eqnarray*}
&&\exp\left( \Omega W \left( \Omega , j \right) \right) = \\
&& \int \mathcal{D} \phi \exp\left\{ -S\left[\phi\right] + 
\Omega j \mathcal{M} \left[ \phi \right] \right\}
\end{eqnarray*}
\begin{eqnarray*}
\mathcal{M}\left[ \phi \right] &=& \frac{1}{\Omega}\int_\Omega \phi\left( x
\right) \mathrm{d}^d x\\
S\left[\phi\right] &=& \int_\Omega \left( \frac{1}{2} \left(
\partial_\mu \phi \right)^2 + V\left[\phi \right] \right) \mathrm{d}^d
x \\
V\left[\phi \right] &=& \frac{1}{2} \mathrm{r}_0 \phi^2 + \frac{1}{4}
\lambda_0 \phi^4.
\end{eqnarray*}

According to ref. \cite{fund2} the CEP $U \left(\Omega,\bar{\phi}
\right)$ is defined as
\begin{eqnarray}
  \label{CEP:definition}
&&  \exp\left( -\Omega U \left(\Omega,\bar{\phi} \right) \right) =\\
&&  \int\mathcal{D} \phi \delta\left( \mathcal{M}\left[ \phi \right]-
    \bar{\phi} \right) \times  \exp\left( - S \left[ \phi \right]
\right) \equiv \mathcal{N} \nonumber
\end{eqnarray}
which entails
\begin{eqnarray}
  \label{CEP:main}
  &&\exp\left( \Omega W \left( \Omega , j \right) \right) = \\ 
&& \int \mathrm{d} \bar{\phi} \exp \left[ \Omega \left( j \bar{\phi} -
U  \left( \Omega, \bar{\phi} \right)
\right) \right] \nonumber
\end{eqnarray}
it has been shown in \cite{fund2} that
\begin{equation}
\label{CEP:lim}
  \lim_{\Omega \to \infty} \Gamma \left( \Omega, \bar\phi \right) =
\lim_{\Omega \to \infty} U \left(\Omega,\bar{\phi} \right).
\end{equation}
\section{Using CEP on the lattice}
If $\Omega$ is large enough, from eq. (\ref{CEP:lim}) we get
\begin{displaymath}
\Gamma \left(\Omega,\bar{\phi} \right) \approx U \left(\Omega,\bar{\phi} \right)
\end{displaymath}
and
\begin{displaymath}
J=\frac{\mathrm{d} \Gamma \left(\Omega,\bar{\phi}
\right)}{\mathrm{d} \bar{\phi}} \approx \frac{\mathrm{d} U
\left(\Omega,\bar{\phi} \right)}{\mathrm{d} \bar{\phi}}.
\end{displaymath}
We define
\begin{eqnarray*}
&&\ACEP{O\left[ \phi \right]} = \\ 
&& \frac{1}{\mathcal{N}} \int\mathcal{D}
\phi \delta\left( \mathcal{M}\left[ \phi \right]- \bar{\phi} \right)
O\left[ \phi \right] \exp\left( - S \left[ \phi \right] \right)
\end{eqnarray*}
From \cite{fund2} CEP gives
\begin{displaymath}
\frac{\mathrm{d} U \left(\Omega,\bar{\phi} \right)}{\mathrm{d}
\bar{\phi}} = \frac{1}{\Omega} \ACEP{ \int \mathrm{d}^d x V'\left[
\phi \right] } = \ACEP{\mathcal{V}'} \approx J
\end{displaymath}
\section{Improving CEP}
We evaluate eq. (\ref{CEP:main}) with the saddle point method around 
a $\varphi$ such that $j-U' \left(\Omega,\varphi \right)=0$. Then we get
\begin{eqnarray*}
  W \left( \Omega , j \right) &=& j \varphi - U \left(\Omega,
    \varphi \right) - \\ && \frac{1}{2} \frac{1}{\Omega} \ln
U''\left(\Omega, \varphi \right) + \\ && \frac{1}{2}
  \frac{1}{\Omega} \ln 2\pi - \frac{1}{2} \frac{ 1}{\Omega} \ln \Omega.
\end{eqnarray*}
From above equation and from eq. (\ref{Gammadef}) then one has
\begin{eqnarray}
  \Gamma \left( \Omega, \varphi \right) &=& U \left(\Omega,  \varphi
\right) + \frac{1}{2} \frac{1}{\Omega} \ln U''\left(\Omega, \varphi
\right) - \label{ICEP:myGamma} \\ && \frac{1}{2} \frac{1}{\Omega} \ln
2\pi + \frac{1}{2} \frac{1}{\Omega}   \ln \Omega \nonumber
\end{eqnarray}
and hence
\begin{equation}
  j = \frac{ \mathrm{d} \Gamma
\left( \Omega, \varphi \right) }{ \mathrm{d} \varphi } = U'
\left(\Omega, \varphi \right) + \frac{1}{2} \frac{1}{\Omega} \frac{
U'''\left(\Omega, \varphi \right) }{ U''\left(\Omega, \varphi \right)
} \label{ICEP:myJ}
\end{equation}
Iteration of the method adopted for computing $U'\left(\Omega, \varphi
\right)$ gives
\begin{eqnarray*}
U''&=& \ACEP{\mathcal{V}''} -\Omega\ACEP{ \left( \mathcal{V}'
-\ACEP{\mathcal{V}'} \right)^2 } \\
U'''&=& \ACEP{\mathcal{V}'''} - \\ && 3 \Omega \ACEP{\left( \mathcal{V}'
-\ACEP{\mathcal{V}'} \right) \left( \mathcal{V}''
-\ACEP{\mathcal{V}''} \right) } + \\
&&\Omega^2 \ACEP{ \left( \mathcal{V}'
-\ACEP{\mathcal{V}'} \right)^3 } 
\end{eqnarray*}
\section{Montecarlo for CEP and ICEP and data analysis}
\Huge
\newsavebox{\LDA}
\sbox{\LDA}{\begin{minipage}[t]{1cm} $\displaystyle \downarrow$ \end{minipage}}
\scriptsize
\newsavebox{\CandI}
\sbox{\CandI}{\begin{minipage}[t]{1cm} \begin{center} CEP \\ ICEP \end{center}
\end{minipage}}

\normalsize
The Montecarlo updating must be performed by keeping constant the
$\bar{\phi}$ value.\par
This can be achieved by doing the Montecarlo update on pairs of
lattice sites in such a way that changing the field values does not
change their average.\par
In \cite{fund2} this was done by choosing a fixed site and pairing it
with the others in turn.\par
We developed a different algorithm, which performs, for each site of
the lattice, a random choice of the second site. This avoids some
problems concerning next neighbor updates and, moreover, allows
encoding either in vectorial or parallel programs.\par
\begin{figure}[tb]
  \begin{center}
    \leavevmode
    \includegraphics[scale=0.9]{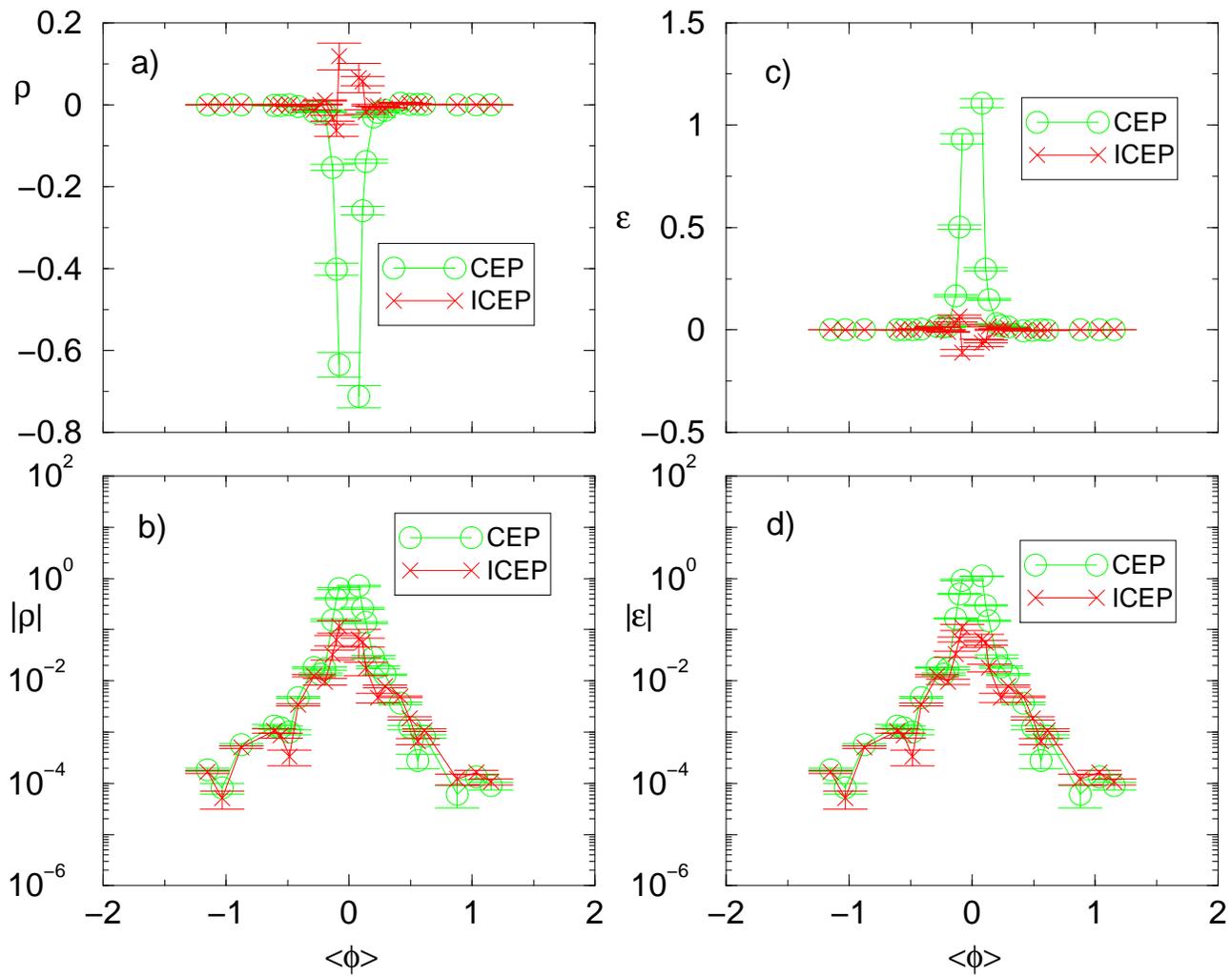}
    \caption{The lattice calculations have been performed with
      $\mathrm{r}_0$=-0.2279 and $\lambda_0$=0.5} 
    \label{fig:cep}
  \end{center}
\end{figure}

Our procedure to get an estimate of finite size effects  
in Montecarlo computations is summarized as follows
\begin{displaymath}
\begin{array}{cccl}
J^\mathrm{IN} & \stackrel{\mathrm{EP}}{\longrightarrow} & \left\langle \phi
\right\rangle^\mathrm{OUT} & \\
&&\usebox{\LDA}&
\begin{array}{l}
\mathrm{mean}\\
\mathrm{value}
\end{array}
\\
J^\mathrm{OUT} & \stackrel{\usebox{\CandI}}{\longleftarrow}& \left\langle \phi
\right\rangle^\mathrm{IN} &
\end{array}
\end{displaymath}
$J^\mathrm{IN}$ is the input value of standard Montecarlo Effective
Potential (EP). In the infinite volume limit $J^\mathrm{IN}$ and
$J^\mathrm{OUT}$, the latter given from CEP or ICEP, should be
equal. Their difference, from finite lattice calculations, 
includes some finite size effects, which should be controlled with
ICEP. The effectiveness of ICEP has been verified from the behavior of
two parameters $ \rho $ and $ \epsilon $
\begin{eqnarray*}
\rho &=& \frac{\left\langle \phi \right\rangle^\mathrm{IN}
J^\mathrm{OUT}}{\left\langle \phi \right\rangle^\mathrm{OUT}
J^\mathrm{IN}} -1 \\
\epsilon &=& 2 \frac{J^\mathrm{IN} - J^\mathrm{OUT}}{J^\mathrm{IN} +
J^\mathrm{OUT}}.
\end{eqnarray*}
The errors in $\rho$ includes those of $\left\langle \phi
\right\rangle^\mathrm{OUT}$ and $J^\mathrm{OUT}$. The
errors in $\epsilon$ come from those in $J^\mathrm{OUT}$ only.
The EP values used as input in our Montecarlo calculations 
are those reported in \cite{myself} supplemented with some others
obtained by running the same program.
They  are also used to compare EP, CEP and ICEP in Fig. \ref{fig:cep}.

\section{Comments and conclusions}
From the plots in Fig. \ref{fig:cep} a similar behaviour of $\rho$
and $\epsilon$ is apparent, though only $\rho$ depends on EP
calculations. The comparison of the usual Montecarlo EP with CEP and
ICEP shows that the latter reduces the finite
size effects, which should also affect EP. These effects are
especially relevant when $\left\langle \phi \right\rangle \approx 0$
and in this domain the plots in Fig. \ref{fig:cep} clearly show that ICEP is better
than CEP.
\section*{Acknowledgements}
The authors are gratefull to Prof. M. Consoli, Prof. P. Cea and
Dr. L. Cosmai for useful discussions and informations in the course of
the work done.

\end{document}